\documentclass[twocolumn,showpacs,preprintnumbers,amsmath,amssymb]{revtex4}
\usepackage{graphicx}
\usepackage[ansinew]{inputenc}

\begin{document}

\renewcommand{\vec}[1]{\ensuremath{\boldsymbol{#1}}}
\newcommand{\Ca}{$^{43}$Ca$^+$}
\newcommand{\Caf}{$^{40}$Ca$^+$}
\newcommand{\bra}[1]{\ensuremath{\langle #1|}}   
\newcommand{\ket}[1]{\ensuremath{| #1\rangle}}   
\newcommand{\qd}{\ket{\mbox{$\downarrow$}}}
\newcommand{\qu}{\ket{\mbox{$\uparrow$}}}

\title{Experimental quantum information processing with \Ca \ ions}

\author{J. Benhelm$^{1,2}$}
\author{G. Kirchmair$^{1,2}$}
\author{C. F. Roos$^{1,2}$}
\author{R. Blatt$^{1,2}$}

\affiliation{$^1$Institut f\"ur Experimentalphysik, Universit\"at Innsbruck, Technikerstr.~25, A-6020 Innsbruck, Austria\\
$^2$Institut f\"ur Quantenoptik und Quanteninformation,
\"Osterreichische Akademie der Wissenschaften, Otto-Hittmair-Platz
1, A-6020 Innsbruck, Austria}

\date{\today}

\begin{abstract}
For quantum information processing (QIP) with trapped ions, the
isotope \Ca \ offers the combined advantages of a quantum memory
with long coherence time, a high fidelity read out and the
possibility of performing two qubit gates on a quadrupole
transition with a narrow-band laser. Compared to other ions used
for quantum computing, \Ca \ has a relatively complicated level
structure. In this paper we discuss how to meet the basic
requirements for QIP and demonstrate ground state cooling, robust
state initialization and efficient read out for the hyperfine
qubit with a single \Ca \ ion. A microwave field and a Raman light
field are used to drive qubit transitions, and the coherence times
for both fields are compared. Phase errors due to interferometric
instabilities in the Raman field generation do not limit the
experiments on a time scale of 100~ms. We find a quantum
information storage time of many seconds for the hyperfine qubit.
\end{abstract}



\maketitle

\section{\label{sec:intro}Introduction}
Quantum information processing with trapped ions has made huge
progress since it was first proposed more than a decade ago
\cite{Cirac1995}. The question of which ion species is best suited
for QIP is still undecided. So far, qubits encoded in trapped ions
come in two flavors. On the one hand, two energy levels of the
hyperfine (or Zeeman) ground state manifold of an ion can serve as
a qubit commonly termed \emph{hyperfine qubit}. The energy
splitting is typically several GHz. Many successful experiments
have been performed on hyperfine qubits with cadmium and beryllium
ions, illustrating the capabilities for QIP
\cite{Leibfried2003b,Leibfried2005,Brickman2005,Haljan2005b}. On
the other hand, quantum information can be encoded in the ground
state and a metastable energy state of an ion. Here the energy
splitting lies in the optical domain and the qubit is therefore
referred to as an \emph{optical qubit}. This concept has been
pursued so far mainly with calcium ions, where all major building
blocks for QIP have been demonstrated
\cite{Schmidt-Kaler2003,Riebe2004,Haeffner2005}.
\par
When it comes to judging the suitability of a particular ion for
QIP, the fidelities and speed of the basic gate operations and the
quantum information storage time are important criteria. Concerning
the latter the hyperfine qubit has been the clearly better choice.
For the optical qubit, it is technically challenging to achieve
quantum information storage times much longer than a few
milliseconds since this requires a laser with a line width of less
than a few Hertz. Yet, optical qubits have excellent initialization
and read out properties. Moreover, the metastable states can be used
as intermediate levels when driving the hyperfine qubits with a
Raman light field as suggested in Ref. \cite{Aolita2007}. Errors
induced by spontaneous scattering, which are a major limitation in
present experiments with Raman light fields detuned from an optical
dipole transition, are then largely suppressed.
\par
From today's perspective, it seems necessary to integrate the
building blocks that have been shown over the years into a single
system. By using hyperfine qubits in combination with metastable
states, it is possible to exploit the best of both concepts. Only a
few ion species offer this possibility, one of them being \Ca. It is
the only calcium isotope with non-zero nuclear spin and it offers
the advantage that all necessary laser wavelengths lying within the
range from 375~nm to 866~nm can be produced by commercially
available solid state lasers. For QIP, only a small number of
electronic levels are of interest (Fig.~\ref{fig:levelscheme}). The
S$_{1/2}$ ground state is split into the states F=4 and F=3 with a
hyperfine splitting of 3.2~GHz \cite{Arbes1994}. It is connected by
electric-dipole transitions to the short-lived levels P$_{1/2}$,
P$_{3/2}$ and by electric-quadrupole transitions to the metastable
states D$_{3/2}$ and D$_{5/2}$ with a lifetime of $\sim$1~s. Because
of the fairly large nuclear spin of $I=7/2$, these levels split into
a total of 144 Zeeman states. This results in a very rich spectrum
for the $S_{1/2}\leftrightarrow D_{5/2}$ transition which has been
investigated to a high precision recently \cite{Benhelm2007}.

\section{Experimental setup}

\Ca \ ions are loaded from an enriched source into a linear Paul
trap by a two step photo-ionization process (423~nm and 375~nm)
\cite{Benhelm2007,Gulde2001,Lucas2004}. Radial confinement is
provided by a quadrupole field created by the application of a radio
frequency voltage to two out of four blade electrodes and connecting
the other two blade electrodes to ground \cite{Schmidt-Kaler2003a}.
Axial confinement is achieved by setting two tip electrodes to
DC-voltages of 500-1500~V, resulting in center-of-mass (com) mode
secular trapping frequencies of $\omega_{ax}/2\pi=0.8-1.5$~MHz.
Applying slightly unequal voltages to the tips leads to an axial
shift in the ions' equilibrium position. The closest distance
between the ions and the tip (blade) electrodes is $2.5$~mm
($0.8$~mm). Two additional electrodes compensate for external
electric stray fields in the radial directions. One of them, located
at a distance of 7.3~mm to the trap center, can also be used to
guide microwave signals to the ions. The trap is housed in a vacuum
environment with a pressure below 2$\cdot$10$^{-11}$~mbar. When no
laser light is present, ion storage times as long as two weeks have
been observed \footnote{With two \Caf \ ions trapped we repeatedly
observe that one of the ions forms a CaOH$^+$ molecule by measuring
the change of the axial sideband frequencies. These events occur
every few hours.}.\par

\begin{figure}[t]
\includegraphics[width=8cm]{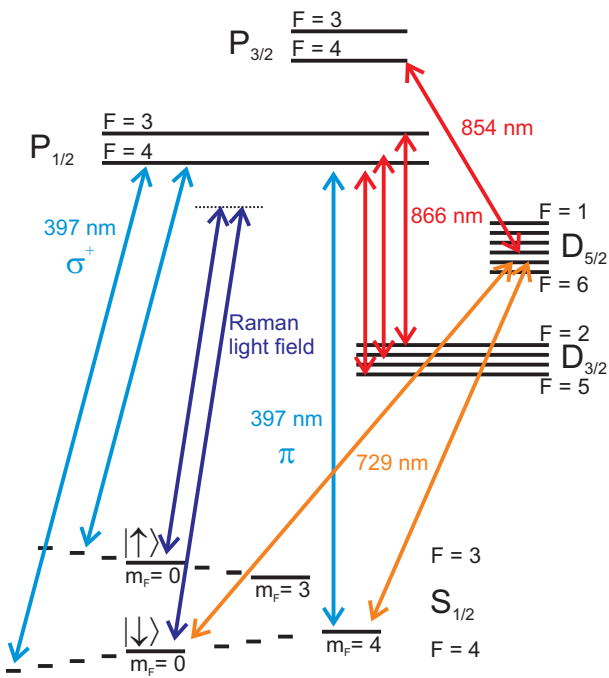}
\caption{\label{fig:levelscheme}  (Color online) Energy level
diagram of the valence electron of \Ca \ showing the hyperfine
splitting of the lowest energy levels. Laser light at 397~nm is used
for Doppler cooling and detection; the lasers at 866 and 854~nm pump
out the metastable D-states. A laser at 729~nm excites the ions on
the transition from the S$_{1/2}$~(F=4)-states to the D$_{5/2}$
(F=2,..,6)-states. It is used for ground state cooling, state
initialization and state discrimination. Microwave radiation applied
to an electrode close to the ions as well as a Raman light field at
397~nm can drive transitions between different levels in the
hyperfine structure of the S$_{1/2}$-state manifold.}
\end{figure}

For most experimental steps, we use a Titanium-Sapphire laser at
729~nm for coupling the energy levels of the S$_{1/2}$(F=4) and
D$_{5/2}$(F=2,..,6) manifolds via their electric-quadrupole
transition (Fig.~\ref{fig:levelscheme}). The laser's frequency is
stabilized to an ultra-stable Fabry-P\'{e}rot cavity
\cite{Notcutt2005} with a line width of 4.7~kHz. A Lorentzian fit to
a beat note measurement with another similar laser reveals a width
of the beat note's power spectral density of 1.8~Hz (4~s data
acquisition time). This is indicative of a line width for each laser
below 1~Hz. In a measurement where a single \Ca \ ion served as a
frequency reference \cite{Benhelm2007} we obtained a line width of
16~Hz with an integration time of 60~s.\par

\begin{figure}[t]
\includegraphics[width=8cm]{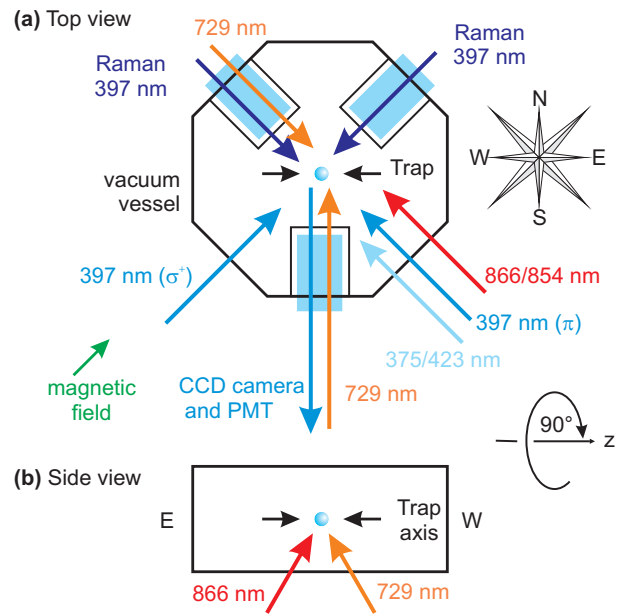}
\caption{\label{fig:opticalaccess} (Color online) (a) Most laser
beams are sent to the ion from a plane containing also the symmetry
axis of the trap (top view). With three custom made lenses in the
inverted viewports S, NE and NW we can tightly focus light at
wavelengths 729~nm and 397~nm. The lens in S is also used to collect
fluorescence light which is sent to a PMT or a camera. The
quantization axis is defined by a magnetic field along SW-NE. (b)
Two beams (729~nm and 866~nm) are sent in from below in a $60°$
angle to the trap axis (side view). The axial trapping potential is
provided by two tips indicated by arrows along the z-axis.
Unbalancing the tip voltages results in a shift  of the ion crystal
along the trap axis (E-W).}
\end{figure}

In order to stabilize the laser's frequency to the atomic transition
frequency, we use a feedback loop based on spectroscopic
measurements typically taken every one to two minutes. Since Ramsey
phase experiments are used to probe the frequencies of two different
Zeeman transitions, these measurements serve also to infer the
strength of the magnetic field at the ion's position. The result is
automatically analyzed and fed back to an acousto-optic modulator
(AOM) located between the laser and its reference cavity. By this
means the laser's output frequency remains constant with respect to
the ion's quadrupole transition frequency. The mean frequency
deviation between laser and ion depends on the Ramsey probe time,
the measurement interval and the actual frequency drift of the
reference cavity and is typically smaller than 200~Hz. With the
knowledge of the laser frequency and the magnetic field, the
transition frequency of all required Zeeman transitions are
calculated to an accuracy of $\pm500~$Hz. The optical frequency and
phase of the laser are controlled with a double-pass AOM (270~MHz)
between laser and ion. Slow drifts of the magnetic field
($\lesssim$200~$\mu$G/h) are taken into account by properly
adjusting the radio frequency feeding the AOM. Deteriorating effects
due to magnetic field noise components at 50~Hz, typically on the
order of 1~mG, are largely suppressed by triggering all experiments
to the AC-line frequency.\par

The hyperfine qubit can be driven with a Raman light field
comprising two phase stable frequency components. It is derived from
a commercial diode laser system, consisting of a tapered amplifier
and a frequency doubling stage emitting at a wavelength of 397~nm.
To achieve the frequency difference of 3.2~GHz required for bridging
the ground state hyperfine splitting in \Ca, the light is first sent
through an AOM (1~GHz) that splits the laser beam into a blue beam
line (+1$^{st}$ order of diffraction) and a red beam line (0$^{th}$
order). The latter passes another AOM operated at 1~GHz (-1$^{st}$
order of diffraction). The remaining frequency shift is achieved by
two more AOMs ($\sim$ 300~MHz) in each beam line. For
non-copropagating Raman light fields, the two beam lines are
separately guided to the ions through the viewports labelled NW and
NE in Fig.~\ref{fig:opticalaccess} (a).\par

Frequency, phase and amplitude control of these lasers is synonymous
to controlling the radio frequency signals applied to the AOMs. We
use a home-made versatile frequency source (VFS) based on direct
digital synthesis that can phase-coherently provide 16 different
radio frequencies up to 305~MHz. Amplitude shaping is achieved with
a variable gain amplifier controlled by a field programmable gate
array. The VFS and all other radio frequency sources providing the
input signals of the AOMs mentioned above are referenced to an
ultra-stable quartz oscillator with a long term stabilization
provided by the global positioning system. For direct microwave
driving the hyperfine qubit, the output of the VFS is mixed with a
signal of 1.35~GHz, then filtered, frequency doubled, filtered and
amplified. In this way, full amplitude, frequency and phase control
of the VFS is up-converted to 3.2~GHz.\par

The laser sources at 866~nm, 854~nm and 397~nm are commercially
available diode lasers whose frequencies are referenced to
Fabry-P\'{e}rot cavities. The Doppler cooling laser at 397~nm is
produced by frequency-doubling light of a near-infrared laser diode.
Except for the Raman light fields, all other light sources are
linked to the experiment with single mode fibers. As sketched in
Fig.~\ref{fig:opticalaccess}, the vacuum vessel provides optical
access for illuminating the ion by laser beams mostly arriving in a
plane containing also the symmetry axis of the trap. In addition,
two beams used for laser cooling (729~nm and 866~nm) are sent in
from below with a $60°$ angle to the trap axis.\par

Fluorescence light is collected with a custom designed lens
($\textnormal{NA}=0.3$) correcting for aberrations induced by the
vacuum window. The light is sent to a photo multiplier (PMT) or a
sensitive camera with a magnification of 25 and a resolution of
2.2~$\mu$m. For a single \Ca \ ion, the signal to noise ratio at the
PMT is typically around 50. The same type of lens is also used to
focus light at 729~nm and 397~nm from the S, NE and NW viewports.

\section{Initializing the $^{43}$C$\mbox{a}^+$ hyperfine qubit}

There are many ways to encode quantum information in the \Ca \ level
structure. An optical qubit with vanishing sensitivity to magnetic
field fluctuation has been proposed in Ref. \cite{Benhelm2007}.
Here, we consider the hyperfine ground state manifold depicted in
Fig.~\ref{fig:levelscheme} where the energy splitting between the
F=3 and F=4 manifold is about 3.2~GHz. For low magnetic fields, the
two states $\qd\equiv$ S$_{1/2} $(F=4, m$_F$=0) and $\qu\equiv$
S$_{1/2} $(F=3, m$_F$=0) exhibit no linear Zeeman effect and are
therefore attractive as a robust quantum information carrier
\cite{Lucas:2007}. As in classical computing, also QIP devices need
to be initialized. In our experiment, the initialization step
comprises Doppler cooling, optical pumping, cooling to the motional
ground state and state transfer to \qd.

\subsection{Doppler cooling and optical pumping}
For Doppler cooling and fluorescence detection, the ion is excited
on the S$_{1/2}$ $\leftrightarrow $P$_{1/2}$ dipole transition with
two laser beams. The beam entering from SE is $\pi$-polarized and is
slightly red detuned from the transition S$_{1/2}$(F=4)
$\leftrightarrow$ P$_{1/2}$(F=4). The second beam is
$\sigma^+$-polarized. It is sent through an electro-optic phase
modulator (3.2~GHz) to excite the ions from the S$_{1/2}$(F=3) and
S$_{1/2}$(F=4) to P$_{1/2}$(F=4) manifold. Coherent population
trapping is avoided by lifting the degeneracy of the Zeeman
sub-levels with a magnetic field. To avoid population trapping in
the D$_{3/2}$ manifold, repumping laser light at 866~nm is applied.
The repumping efficiency was improved by tuning the laser close to
the D$_{3/2}$(F=3)$\leftrightarrow$ P$_{1/2}$(F=3) transition
frequency and providing two additional frequencies shifted by
-150~MHz and -395~MHz such that all hyperfine $D_{3/2}$-levels are
resonantly coupled to one of the P$_{1/2}$(F=3,4)-levels. We
observed a maximum fluorescence count rate of 24~kcps per \Ca \ ion
on a PMT for magnetic fields ranging from 0.2 to 5~G. This is about
45\% of the count rate we observe for \Caf \ ions. The count rate
difference, possibly caused by coherent population trapping, is
still under investigation.\par

\begin{figure}[t]
\includegraphics[width=8cm]{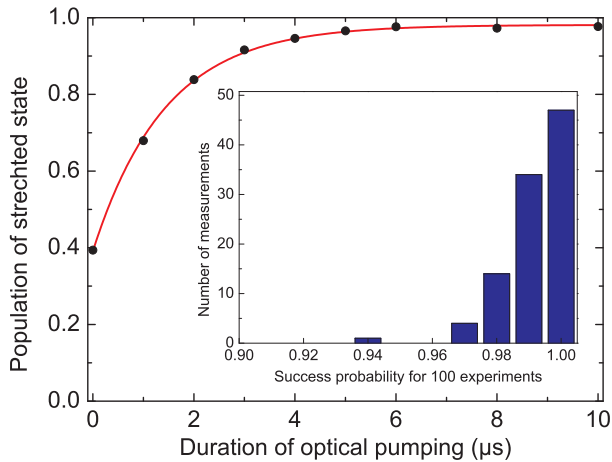}
\caption{\label{fig:opticalpumping} (Color online) The population in
the stretched state S$_{1/2}$(F=4, m$_F$=4) is plotted as a function
of the duration of optical pumping. An exponential fit (solid line)
reveals a time constant of 1.4~$\mu$s. After 10~$\mu$s the
population is in the desired state in 98~\% of the measurements. The
inset shows a histogram of the success rate of 100 measurements each
containing 100 experiments when two $\pi$-pulses on the quadrupole
transition are applied and an additional intermediated optical
pumping interval is used. This enhances the fidelity of the process
to above 99.2\%.}
\end{figure}

After switching off the $\pi$-polarized laser beam, the ion is
optically pumped into the state S$_{1/2}$(F=4, m$_F$=4). The state's
population was measured with two consecutive $\pi$-pulses exciting
the population to the D$_{5/2}$-state and subsequent fluorescence
detection (section \ref{sec:statedetection}).
Figure~\ref{fig:opticalpumping} shows the dynamics of optical
pumping and illustrates that the stretched Zeeman states of the
ground state manifold are already strongly populated during
Doppler-cooling. An exponential fit to the data points yields a time
constant of the process of 1.4~$\mu$s. After 10~$\mu$s, the desired
state is populated in $98$\% of the cases.\par

The pumping efficiency can be improved by transferring the
population after this first step with a $\pi$-pulse to the D$_{5/2}$
(F=6, m$_F$=6)-state and repeating the optical pumping. By applying
another $\pi$-pulse on the same transition, the populations in
S$_{1/2}$(F=4, m$_F$=4) and D$_{5/2}$(F=6, m$_F$=6) are exchanged.
On average 98\% should now be in S$_{1/2}$(F=4, m$_F$=4) and the
rest in the D$_{5/2}$-state. Finally the two populations are
combined by switching on the 854~nm laser for a short time to clear
out the D$_{5/2}$-state via the P$_{3/2}$ (F=5, m$_F$=5)-state from
where it can decay only into the desired stretched state. The inset
of Fig.~\ref{fig:opticalpumping} shows a histogram built from 100
measurements each comprising 100 experiments indicating a lower
bound of the pumping efficiency of 99.2\%.\par

After Doppler cooling and optical pumping, an average population
$\bar{n}_{ax}=10(5)$ of the axial mode is inferred from measuring
the decay of Rabi oscillations on the blue axial sideband. The
average number of quanta is heavily dependent on the different laser
detunings and powers.

\subsection{Ground state cooling, heating rate and ion shuttling}

\begin{figure}[t]
\includegraphics[width=8cm]{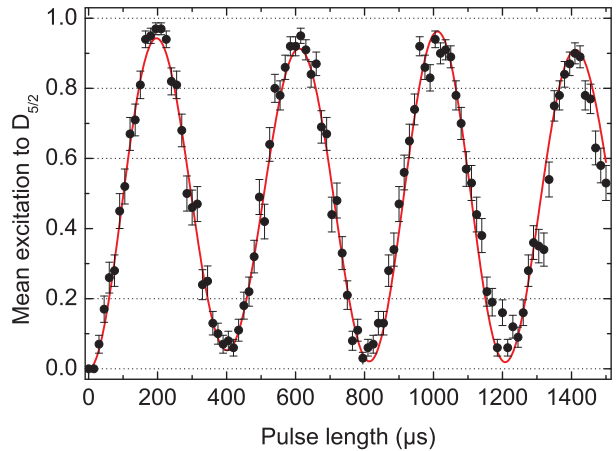}
\caption{\label{fig:blueflops} (Color online) Rabi oscillations on
the blue axial sideband of the transition S$_{1/2}$(F=4,
m$_F$=4)$\leftrightarrow$ D$_{5/2}$(F=6, m$_F$=6) after ground state
cooling. The solid line is a fit assuming a thermal state. It yields
a mean occupation of the axial mode of $\bar{n}_{ax}=0.06$.}
\end{figure}

Cooling the ions to the motional ground state is mandatory in order
to maximize quantum gate fidelities. In our experiment, it has been
implemented with a scheme analogous to what has been demonstrated
with \Caf \ ions \cite{Roos1999}. In order to obtain a closed
cooling cycle, the frequency of the laser at 729~nm is tuned to the
red sideband ($\omega_{ax}/2\pi=1.18$~MHz) of the transition
S$_{1/2}$(F=4, m$_F$=4)$\leftrightarrow$ D$_{5/2}$(F=6, m$_F$=6). An
additional quenching laser at 854~nm is required to increase the
spontaneous decay rate to the energy level S$_{1/2}$ by coupling the
D$_{5/2}$(F=6, m$_F$=6) to the P$_{3/2}$ (F=5, m$_F$=5)-state.
Spontaneous decay to the stretched state is taking the entropy away
from the ion. In each cycle, one motional quantum can be removed.
The residual occupation of the motional mode is measured by
comparison of the red and the blue sideband excitation.
Alternatively, Rabi oscillations on a blue motional sideband can be
observed in order to infer the average population of the axial mode
(see Fig.~\ref{fig:blueflops}). The solid line is a fitted model
function with $\bar{n}_{ax}$ as a free parameter. From both methods,
we consistently obtain $\bar{n}_{ax}=0.06$.\par

By introducing and varying a delay time between ground state cooling
and the temperature measurement, we determined a heating rate on the
axial com-mode of 1 motional quantum per 370~ms. The coherence of a
motional superposition state $|0\rangle +|1\rangle$ was investigated
by performing Ramsey experiments that mapped the motional
superposition states after a variable waiting time to the $S_{1/2}$
and $D_{5/2}$ electronic states. These measurements showed that the
motional coherence was preserved for more than 320(10)~ms, in good
agreement with the measured heating rate.\par

When it comes to scaling the system up to strings of many ions, it
is important that single qubit gates can be applied to each
individual ion. Individual addressing can be achieved by using an
electro-optical deflector to rapidly steer a strongly focussed laser
beam to different ions in the string with high precision
\cite{Schmidt-Kaler2003a}. Driving Raman transitions in \Ca \
requires more than a single laser beam that would have to be steered
this way. To avoid this complication, we prefer to shuttle the ion
string along the axis of the trap instead of moving the laser beams.
For an axial trapping frequency of $\omega_{ax}/(2\pi)=1.18$~MHz we
are able to shuttle the ions over a distance of up to 10~$\mu$m by
changing the right (left) tip voltage from 990~V(1010~V) to
1010~V(990~V). The switching speed is currently limited by a low
pass filter with a cutoff at 125~kHz that prevents external
electrical noise from coupling to the trap electrodes. Shuttling
over the full distance in order to individually address single ions
works for transport durations as low as 40~$\mu$s. In a test run
with a single \Caf \ ion, quantum information encoded into the
motional states \ket{n=0} and \ket{n=1}, was fully preserved during
the shuttling.\par

\subsection{Transfer to the hyperfine clock states}

Ground state cooling on quadrupole transitions requires a closed
cooling cycle which can only be achieved efficiently when working
with the stretched hyperfine ground states (F=4, m$_F$=$\pm$4).
For this reason, methods are needed that allow for a transfer from
these states to the qubit state \qd. Four different techniques
were under consideration:

\subsubsection{Optical pumping on the S$_{1/2}$ to P$_{1/2}$ transition}

The state \qd \ could be populated by optical pumping with
$\pi$-polarized light fields exciting the transitions S$_{1/2}$(F=4)
$\leftrightarrow$ P$_{1/2}$ (F=4) and S$_{1/2}$ (F=3)
$\leftrightarrow$ P$_{1/2}$ (F=4) within a few microseconds.
However, many scattering events would be required to pump the
population to the desired state that are likely to heat up the ion
from the motional ground state. Moreover, the efficiency of the
optical pumping would probably be fairly poor as small polarization
imperfections of the beams and repumping via the S$_{1/2}$(F=4)
$\leftrightarrow$ P$_{1/2}$(F=3) are likely to occur.

\subsubsection{Raman light field}
Transferring the population can also be achieved with a Raman light
field detuned from the S$_{1/2} \leftrightarrow$ P$_{1/2}$ dipole
transition at 397~nm. In the simplest scenario, a sequence of four
$\pi$-pulses would be used to populate the state \qd \ starting from
S$_{1/2}$(F=4, m$_F$=$\pm$4) by changing the magnetic quantum number
in units of $\Delta m=\pm1$. Use of copropagating beams suppresses
unwanted excitations of motional sidebands.

\subsubsection{Microwave}

Instead of a Raman light field, also a microwave field can be used
to transfer the ions in a four-step process to \qd. An additional
advantage here is that the field's wavelength is huge compared to
the distance of the ions and therefore an equal coupling of all ions
to the field is guaranteed.

A limitation for both methods, Raman light field and microwave, is
the small coupling strength on the transitions (F=3, m$_F$=$\pm$ 3)
$\leftrightarrow$ (F=4, m$_F$=$\pm$2). That makes the whole process
either slow or necessitates a larger frequency separation of the
Zeeman levels in order to suppress non-resonant excitation of
neighboring transitions.

\begin{figure}[t]
\includegraphics[width=8cm]{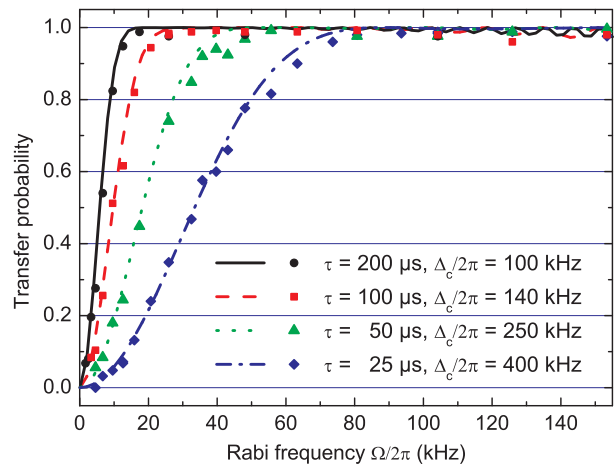}
\caption{\label{fig:rap} (Color online) Transfer probability
measurement of an amplitude shaped laser pulse on the transition
S$_{1/2}$ (m$_F$=1/2) $\leftrightarrow$ D$_{5/2}$(m$_F$=3/2) of a
single \Caf \ ion as a function of the Rabi frequency. Data were
taken for four different pulse lengths $\tau$ and frequency chirp
spans $\Delta_c$ as given in the plot legend. The lines indicate
what is theoretically expected. With enough laser power available
the transfer probability hardly changes over a wide range of Rabi
frequencies.}
\end{figure}

\begin{figure*}[t]
\includegraphics[width=16cm]{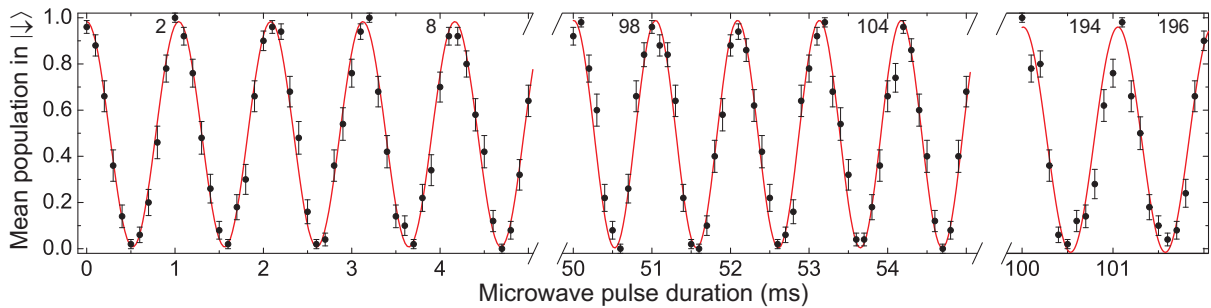}
\caption{\label{fig:mwflops} (Color online) Rabi oscillations on the
\Ca \ hyperfine qubit mediated by a microwave field after 0, 50 and
100~ms. Each data point represent 50 individual measurements. The
solid line is a weighted least square fit with the function
$\frac{A}{2}\cos(\pi\cdot t/\tau_\pi)+y_0$ resulting in $y_0=
0.490(3)$, $\tau_\pi=520.83(3)~\mu$s and $A=0.974(11)$. The number
of state transfers are indicated. Since the amplitude of the Rabi
oscillation is still close to unity even after 200 state transfers
the microwave can serve as a reference to the Raman light field
regarding power and phase stability.}
\end{figure*}

\subsubsection{Transfer via quadrupole transitions}
State transfer based on a laser operating on the quadrupole
transition S$_{1/2} \leftrightarrow$ D$_{5/2}$ reduces the transfer
process to two steps since the selection rules allow for $\Delta
m=\pm2$. The duration of a $\pi$-pulse can be as short as a few
microseconds, and only a single laser beam is needed that can be
either focused to a small region or illuminate the whole trap
volume. If the D$_{5/2}$(F=4) is chosen as intermediate state, a
good compromise is achieved between the quadrupole coupling strength
of the involved transitions and the frequency separation of the
neighboring D-state Zeeman levels. The latter is by a factor 1.6
larger as for the ground states. In particular for low magnetic
fields this method is expected to work better than a transfer with
Raman or microwave fields.
\par
With the precision laser for the quadrupole transition acting on a
single \Ca \ ion, the implementation of state transfer is
straightforward. With two consecutive $\pi$-pulses we achieved a
transfer success probability of more than 99\%. Assuming Gaussian
beam waists, such high probabilities cannot be expected for larger
ions crystals though, unless one is willing to waste most of the
laser power by making the beam size very large. As variations of the
coupling strengths may also arise from other technical
imperfections, a more robust scheme seems to be desirable. Inspired
by Ref. \cite{lu2005}, we introduce amplitude shaping and a linear
frequency sweep of the transfer pulses to demonstrate a transfer
technique less sensitive to changes in the laser intensity.
Figure~\ref{fig:rap} shows the transfer probability for a single
\Caf \ ion and the transition S$_{1/2}$(m=1/2) $\leftrightarrow$
D$_{5/2}$(m=3/2) as a function of the Rabi frequency for four
different pulse durations $\tau$. The amplitude of the laser pulse
had a $\cos^2$-shape over the pulse length. The frequency of the
laser was linearly swept over a range $\Delta_c$ centered on the
transition frequency. The data show clearly that the transfer
probability is hardly affected over a broad range of Rabi
frequencies $\Omega$ for the different parameters.

\section{State detection}
\label{sec:statedetection}

For \Ca \ ions, the electron shelving technique first introduced by
Dehmelt allows for an efficient state discrimination between the \qd
\ and \qu \ hyperfine qubit states by scattering light on the
S$_{1/2}$ to P$_{1/2}$ transition after having shelved the \qd \
state in the D$_{5/2}$ metastable state with a $\pi$-pulse. In our
experiment, the same light fields as for Doppler cooling are used
but with slightly more power. With this method, not only \qu \ and
\qd \ can be discriminated but the other Zeeman levels in the
S$_{1/2}$ and D$_{5/2}$-state manifolds, too. The quality of the
transfer pulses sets a limitation to the state discrimination.
Again, pulse shaping and frequency sweeping can help to increase the
robustness with respect to intensity variations of the shelving
laser. In addition, instead of using a single $\pi$-pulse excitation
to a certain Zeeman state in the D$_{5/2}$-state manifold, the first
$\pi$-pulse can be followed by a second one, exciting any population
still remaining in \qd \ to a different Zeeman state. Assuming a
transfer probability of 0.99 for each of the pulses, one expects a
transfer error probability of less then 10$^{-4}$. The final
detection fidelity will then be limited by spontaneous decay from
the D$_{5/2}$-state during the detection whose duration depends on
the signal-to-noise ratio and signal strength. For the experiments
reported here, the detection time was set to 5~ms. The error due to
spontaneous decay is estimated to be 0.5$\%$.

\section{Single qubit gates on the $^{43}$C$\mbox{a}^+$ hyperfine qubit}

Once external and internal degrees of freedom are initialized,
quantum information needs to be encoded into the ions, stored and
manipulated. This is achieved with a driving field tuned to the
qubits' transition frequency. Two different driving fields were
investigated.

\subsection{Microwave drive}

From an experimental point of view, quantum state manipulation by
microwave radiation is simple and robust. There is no alignment
required and stable frequency sources are readily available with
computer-controlled power, frequency and phase.
\par
To characterize the microwave properties on a single \Ca \ ion, Rabi
oscillations were recorded on the hyperfine qubit \qd \
$\leftrightarrow$ \qu \ at a magnetic field of 3.4~G. After
initializing the ion into \qd \ a microwave signal of 3.226~GHz is
turned on for a variable amount of time followed by state detection.
Figure~\ref{fig:mwflops} shows the resulting Rabi oscillations at
instances of 0, 50 and 100~ms. The solid line represents a weighted
least square fit with the function $f(t)=\frac{A}{2}\cos(\pi\cdot
t/\tau_\pi)+y_0$ resulting $y_0=0.490(3)$, $\tau_\pi=520.83(3)~\mu$s
and $A=0.974(11)$. About 200 state transfers are observed over a
time of 100~ms with hardly any decrease in fringe amplitude. Also
for measurements with $\tau_\pi= 34.3~\mu$s a fringe amplitude A
close to unity has been observed for more than 150 state transfers.
In both cases, the subsequent decay of the fringe amplitude for more
oscillations indicates a limitation due to small fluctuations of the
microwave power.
\par
Unfortunately, microwave excitation does not couple motional and
electronic states unless strong magnetic field gradients are applied
\cite{Mintert2001} and it cannot be focussed to a single qubit
location. Nevertheless, microwave excitation turns out to be a
useful reference for investigating the phase stability of Raman
excitation schemes to be discussed in the next paragraph.

\subsection{Raman light field}

\begin{figure}[t]
\includegraphics[width=8cm]{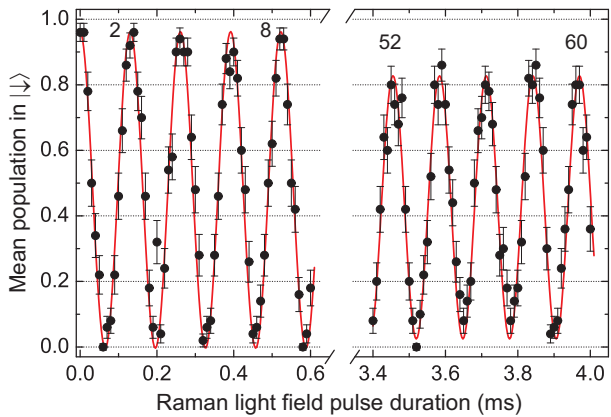}
\caption{\label{fig:ramanflops} (Color online) Rabi oscillations on
the \Ca \ hyperfine qubit induced by a colinear Raman light field.
Each data point represents 50 individual measurements. As for the
microwave excitation, we fitted a sinusoidal function to the data
set from 0 to 600~$\mu$s which yields a fringe amplitude
$A=0.97(1)$, a $\pi$-time of $\tau_\pi=65.3(1)~\mu$s and a fringe
center at $y_0=0.479(5)$. Fitting to the data points beyond 3.4~ms a
small offset phase had to be introduced and $\tau_\pi$ adjusted to
63.8(2)~$\mu$s indicating a small increase of the Raman light power
during the measurement. The Amplitude reduces to $A=0.80(2)$ and the
fringe center dropped to $y_0=0.428(7)$. A comparison with microwave
excitation reveals imperfections caused by spontaneous scattering
and laser amplitude fluctuations.}
\end{figure}

In contrast to the microwave drive, the interaction region of the
Raman field detuned from the dipole transition S$_{1/2}$
$\leftrightarrow$ P$_{1/2}$ is as small as the diameter of the
involved laser beams. The coupling to the motional mode along the
trap axis (unit vector $\vec{e}_z$) is described by the Lamb-Dicke
parameter $\eta$=$(\vec{k}_+-\vec{k}_-)\cdot
\vec{e}_z\sqrt{\frac{\hbar}{2 M \omega}}$ where \vec{k}$_\pm$ is the
wave vector of the blue and the red Raman light field respectively,
M the atomic mass and $\omega$ denotes the trap frequency. For
copropagating lasers the Lamb-Dicke factor is negligible whereas it
is maximized for lasers counter-propagating along the motional mode
axis.\par

We characterize the Raman interaction on a single ion by driving
Rabi oscillations on the hyperfine qubit with copropagating beams
from NW that are detuned from the S$_{1/2}$ to P$_{1/2}$ transition
frequency by -10~GHz. Figure~\ref{fig:ramanflops} shows Rabi
oscillations for excitation times of up to 4~ms with a duration of a
$\pi$-pulse of $\tau_\pi=65.3(1)~\mu$s. The first few oscillations
have a fringe amplitude of $A=0.97(1)$ which is reduced to $0.80(2)$
after more than 50 state transfers. Shot to shot variations in Raman
light intensity contribute to a loss symmetrically to the average
excitation. In addition the fringe center, ideally at $y_0=0.5$, has
dropped to $y_0=0.428(7)$ due to non-resonant scattering introduced
by the Raman light field.\par

The ability to couple electronic and motional states by the Raman
excitation was tested by comparing Rabi frequencies on the carrier
and on the first blue sideband with non-copropagating beams (from NW
and NE) illuminating an ion initially prepared in the motional
ground state. The two Raman beams enclose a 90$°$ angle such that
the residual momentum transfer is optimized for the axial direction.
From the ratio of the Rabi frequencies, we directly infer the
Lamb-Dicke parameter to be $\eta=0.216(2)$ in good agreement with
the theory.

\begin{figure}[t]
\centering
\includegraphics[width=7.25cm]{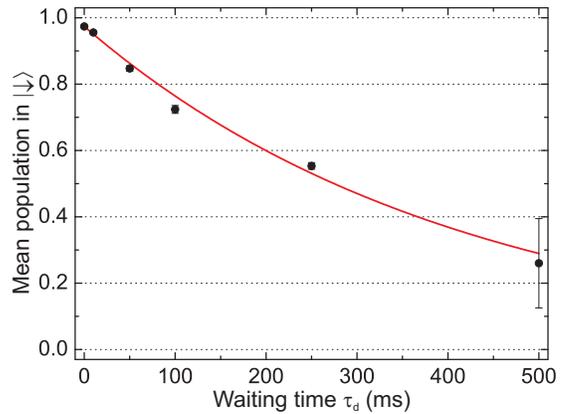}
\caption{\label{fig:bluescatter} (Color online) Measurement of the
qubit state \qd- population probability after a waiting time
$\tau_d$. Single photon scattering events induced by residual light
at 397~nm lead to a transfer of population from the \qd -state to
other Zeeman states in the ground state manifold. The solid line is
an exponential fit with a decay time constant of 410~ms.}
\end{figure}

\section{Coherence properties of the $^{43}$C$\mbox{a}^+$ hyperfine qubit}

\begin{figure}[t]
\includegraphics[width=8.5cm]{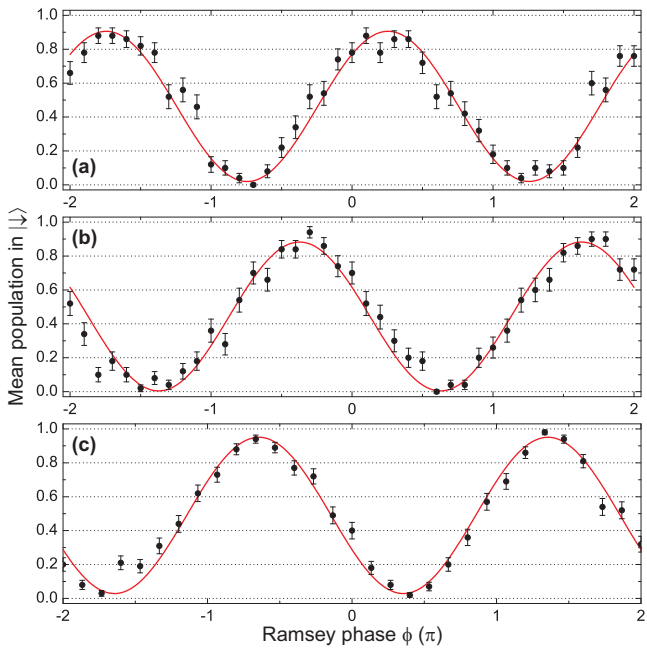}
\caption{\label{fig:raman_coherence} (Color online) Ramsey phase
experiments on the \Ca \ hyperfine qubit at a magnetic field of
3.4~G with a Ramsey waiting time $\tau_R$ set to 100~ms. The data
were taken with three different driving fields. (a) Microwave drive
with $\tau_\pi=19~\mu$s, (b) copropagating Raman light field with
$\tau_\pi=20~\mu$s and (c) non-copropagating Raman light field where
$\tau_\pi=23~\mu$s. The fringe amplitudes are determined by weighted
least square sinusodial fits with amplitude A, offset phase $\Phi_0$
and fringe center $y_0$ as free parameters. This yields fringe
amplitudes of 0.886(17), 0.879(16) and 0.922(13) respectively.
Dephasing by interferometric instabilities is not limiting the
experiments on these timescales. For Ramsey times beyond 100~ms we
observed a further decay of the fringe amplitude.}
\end{figure}

Applying the methods described before, we investigated quantum
information storage capabilities of the \Ca \ hyperfine qubit.
Limitations to the coherence time arise from both spontaneous
scattering events and dephasing \cite{Lucas:2007}. For the hyperfine
qubit, spontaneous decay is negligible since the lifetime of the
involved states can be considered as infinite for all practical
purposes. Scattering can be induced though by imperfectly switched
off laser beams. To judge the importance of this effect, we prepared
the ion in the state \qd. After waiting for a time $\tau_d$, we
transferred the population with two subsequent $\pi$-pulses to
D$_{5/2}$(F=6, m$_F$=0) and D$_{5/2}$(F=4, m$_F$=2). Ideally no
fluorescence should be observed. Figure~\ref{fig:bluescatter} shows
how an initial \qd-population of 0.97 decreases with increasing
waiting time $\tau_d$. An exponential decay fit yields a time
constant of 410~ms. This observation can be explained by imperfect
switching off the cooling laser at a wavelength of 397~nm by one
single-pass AOM only. For every blue photon that is scattered the
ion will be lost from the state \qd \ with a high probability by
decaying to one of the other S$_{1/2}$ Zeeman states. This
complication was avoided by using a mechanical shutter completely
switching off the Doppler cooling laser in all Rabi and Ramsey
experiments lasting for 50~ms and longer.\par

\begin{figure}[t]
\includegraphics[width=8.5cm]{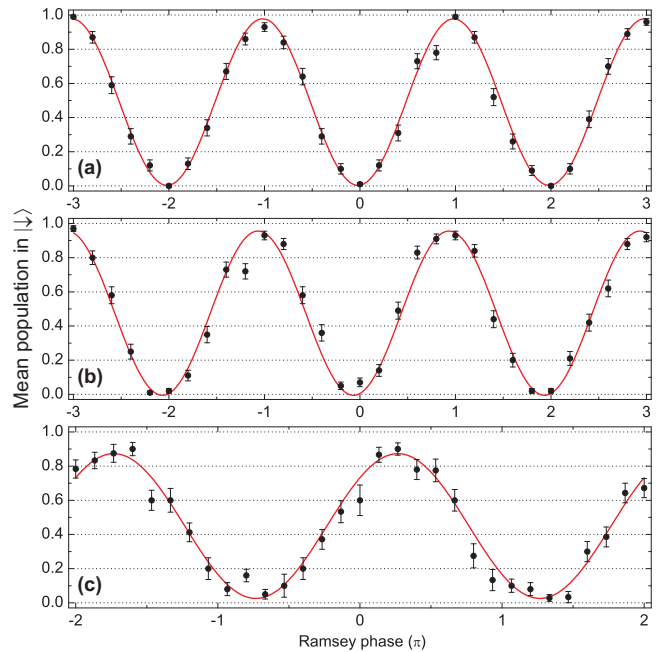}
\caption{\label{fig:mw_coherence} (Color online) Ramsey phase
experiments on the \Ca \ hyperfine qubit with microwave excitation
at a magnetic field of 0.5~G. (a) A scan with $\tau_R=50~\mu$s
results an amplitude of 0.976(4) and demonstrates the ability of
state initialization, manipulation and readout. (b) For a Ramsey
time $\tau_R=200$~ms the amplitude is 0.962(11). (c) For a Ramsey
time of $\tau_R=1$~s a fringe amplitude of 0.847(21) was measured.
Here the measurement time was about 90~min. The reduction of the
amplitude is attributed to the residual sensitivity of 1.2~Hz/mG to
ambient magnetic field fluctuations. }
\end{figure}

Decoherence due to dephasing does not alter the state occupation
probabilities. Instead, the phase information between driving field
and the qubit gets lost. A powerful method to characterize this
effect consists in measuring fringe amplitude in Ramsey phase
experiments. Here a superposition of the two qubit states is created
by a $\pi/2$-pulse. After a waiting time $\tau_{R}$, during which
the qubit evolves freely, a second $(\pi/2)_{\phi}$-pulse is
applied. By scanning the Ramsey phase $\phi$ of the second pulse, a
sinusoidal fringe pattern is observed whose fringe amplitude is a
measure of the coherence.\par

For the Raman light field, the relevant phase is not only determined
by the radio frequency devices supplying the AOMs creating the
3.2~GHz splitting but also by the relative optical path length of
the red and the blue beam line. In general, the absolute phase is
not of interest as long as it does not change during the experiment.
The setup can be considered as an interferometer whose sensitivity
is also dependent on its size. In case of a copropagating Raman
light field, the two beam lines are recombined on a polarizing beam
splitter directly after the relative frequency generation. Here the
interferometer encloses an area of about 0.04~m$^2$. Whereas the
non-copropagating beams enclosing an area of about 0.15~m$^2$. In
order to see whether the experiment would be limited by this effect,
we investigated three different configurations.\par

Figure~\ref{fig:raman_coherence} shows the resulting Ramsey fringe
patterns when driving the hyperfine qubit with a microwave (a), a
copropagating Raman field (b) and a non-copropagating Raman field
(c). The Ramsey waiting time $\tau_R$ was set to 100~ms, the
$\pi/2$-pulses having a duration of about 20~$\mu$s. Each data point
represents either 50 or 100 measurements. The error bars indicate
statistical errors and are used as weights when fitting the function
$f(\phi)=\frac{A}{2}\sin(\phi+\phi_0)+y_0$ to the data in order to
determine the fringe amplitude $A$. The parameters $y_0$ and
$\phi_0$ are also free fit parameters but not further considered.
For the different excitation schemes, we find fringe amplitudes of
0.886(17), 0.879(16) and 0.922(13) respectively. From this we
conclude that errors introduced by interferometric instabilities in
generating the Raman beams do not limit our experiment on time
scales up to 100~ms. For longer Ramsey times, we observed a further
decay of fringe amplitude which we attribute to dephasing. These
measurements were performed at a magnetic field of 3.4~G. For small
magnetic fields, the residual qubit sensitivity to magnetic field
fluctuations increases linearly with a slope of 2.4~kHz/G$^2$.
Therefore we reduced the magnetic field to 0.5~G and repeated the
measurement with the microwave field. The resulting fringe patterns
for three different Ramsey times $\tau_R$ are depicted in Fig.
\ref{fig:mw_coherence}. A short waiting time of $\tau_R=50~\mu$s
results in a fringe pattern amplitude of 0.976(4) (a). This
demonstrates the ability of reliable state initialization, read out
and single qubit gate operation for the \Ca \ hyperfine qubit at low
magnetic fields. For a Ramsey time of $\tau_R=200$~ms we still
obtain a fringe amplitude of 0.962(11) (b). A drop in amplitude to
0.847(21) is observed only after increasing the Ramsey waiting time
to $\tau_R=1$~s (c).\par

Typically, the coherence time is defined as the Ramsey time for
which the fringe amplitude $A$ has dropped to a value of 1/$e$.
Extrapolation of our measurements would lead to a coherence time on
the order of about $6.0$~s assuming an exponential decay and $2.5$~s
for Gaussian decay. Comparing the measurements at $3.4$~G and
$0.5$~G, we conclude that the main limitation to the coherence time
comes from the residual sensitivity of the qubit at finite fields.
Further improvements can be made by means of active magnetic field
stabilization and passive shielding. In addition, rephasing can be
achieved by an intermediate spin-echo pulse that exchanges the
populations of the two qubit levels.
\section{Summary and discussion}
In conclusion, we have discussed and demonstrated various
experimental techniques for high fidelity QIP with \Ca \ ions. These
techniques were applied for measuring the quantum information
storage capabilities of the hyperfine qubit in a noisy environment
to be many seconds. Furthermore, we demonstrated that
interferometric instabilities due to Raman frequency creation do not
limit the phase coherence on time scales up to 100~ms. For most
experimental steps the usage of the quadrupole transition laser is
crucial for our scheme. It seems straight forward to apply these
techniques to strings of ions without compromising the error rate.
From other experiments with \Caf \ ions, we already have
experimental evidence that high fidelity two qubit operations are
possible for the optical qubits \cite{Benhelm2008}. It will be
interesting to explore how these can be combined with the long
storage times found here by swapping quantum information between
hyperfine and optical qubits.
\begin{acknowledgments}
We gratefully acknowledge the support of the European network SCALA,
the Institut für Quanteninformationsverarbeitung and iARPA.
\end{acknowledgments}
\bibliography{apssamp}

\end{document}